\documentclass[english,aps,prl,twocolumn]{revtex4}
\usepackage[T1]{fontenc}
\usepackage[latin9]{inputenc}
\usepackage{color}
\usepackage{babel}
\usepackage{bm}
\usepackage{amsmath}
\usepackage{amssymb}
\usepackage{graphicx}
\usepackage[unicode=true,
 bookmarks=true,bookmarksnumbered=false,bookmarksopen=false,
 breaklinks=false,pdfborder={0 0 0},pdfborderstyle={},backref=false,colorlinks=true]
 {hyperref}

\makeatletter
\@ifundefined{textcolor}{}
{%
 \definecolor{BLACK}{gray}{0}
 \definecolor{WHITE}{gray}{1}
 \definecolor{RED}{rgb}{1,0,0}
 \definecolor{GREEN}{rgb}{0,1,0}
 \definecolor{BLUE}{rgb}{0,0,1}
 \definecolor{CYAN}{cmyk}{1,0,0,0}
 \definecolor{MAGENTA}{cmyk}{0,1,0,0}
 \definecolor{YELLOW}{cmyk}{0,0,1,0}
}

\makeatother

\begin{document}
\title{Magnetization Dynamics in 1D Chains of Ferromagnetic Nanoparticles
Coupled with Dipolar Interactions: Blocking Temperature}
\author{F. Vernay}
\email{francois.vernay@univ-perp.fr}

\author{H. Kachkachi}
\email{hamid.kachkachi@univ-perp.fr}

\date{\today}
\affiliation{Laboratoire PROMES CNRS (UPR-8521) \& Université de Perpignan Via
Domitia, Rambla de la thermodynamique, Tecnosud, F-66100 Perpignan,
France}
\begin{abstract}
There is so far no clear-cut experimental analysis that can determine
whether dipole-dipole interactions enhance or reduce the blocking
temperature $T_{B}$ of nanoparticle assemblies. It seems that the
samples play a central role in the problem and therefore, their geometry
should most likely be the key factor in this issue. Yet, in a previous
work, Jönsson and Garcia-Palacios did investigate theoretically this
problem in a weak-interaction limit and without the presence of an
external DC field. Based on symmetry arguments they reached the conclusion
that the variation of the relaxation rate is monotonous. In the presence
of an external magnetic field we show that these arguments may no
longer hold depending on the experimental geometry. Therefore, the
aim of this paper is to evaluate the variation of $T_{B}$ for a model
system consisting of a chain of ferromagnetic nanoparticles coupled
with long-range dipolar interaction with two different geometries.
Rather than addressing a quantitative analysis, we focus on the qualitative
variation of $T_{B}$ as a function of the interparticle distance
$a$ and of the external field $h$. The two following situations
are investigated: a linear chain with a longitudinal axial anisotropy
in a longitudinal DC field and a linear chain with a longitudinal
axial anisotropy in a transverse field.
\end{abstract}
\maketitle

\section{Introduction}

The magnetization dynamics in magnetic nanoparticles (MNP) assemblies
is the corner stone of many physical observables such as the dynamical
susceptibility, the magnetic resonance and many others. However, having
a direct access to these observables does not guarantee that one is
able to disentangle the collective effects from the intrinsic properties
of the isolated nanoparticles. The competition between these two effects
might indeed impair the true picture that one should have of the long
range interaction physics in such complex systems \citep{sabsabietal13prb,Vernay_etal_acsucept_PRB2014}.
For this reason, and thanks to the long-standing endeavor that has
been devoted to the study of MNP assemblies, it is desirable to first
deal with the relatively simple and ordered low-dimensional systems.
Quasi bi-dimensional assemblies are nowadays well controlled by chemists,
see \emph{e.g.} \citep{Toulemon_etal_2d_2011,Pauly_Pichon_etal_2dassembly_JMatChem_2011,Pauly_Pichon_etal_2dassembly_JMatChem_2012},
and 1D chains of magnetic nanoparticles have been investigated for
more than 30 years in magnetotactic bacteria \citep{mann1985structure}
with a recent revival of interest \citep{bazylinski2004magnetosome,Charilaou_FMR_chains_JAP2011,myrovali2016arrangement}
due to their potential applications. In order to clearly highlight
the effect of dipole-dipole interactions (DDI) on the dynamics of
MNP assembly, we tackle the problem by determining semi-analytically
the behavior of the blocking temperature $T_{B}$ in 1D chains of
MNP, which requires the calculation of the relaxation rate of the
chain.

\begin{figure}
\begin{centering}
\includegraphics[width=0.8\columnwidth]{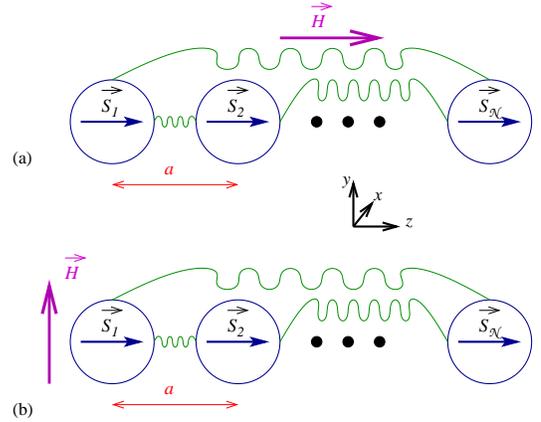}
\par\end{centering}
\caption{\label{fig:1D-chain}1D chain of $\mathcal{N}$ magnetic nanoparticles.
All magnetic nanoparticles are interacting via long range dipolar
interaction depicted in green, each nanoparticle has a uniaxial anisotropy
axis along $z$. Two situations are considered: (a) a longitudinal
field with respect to the chain and to the anisotropy axis, or (b)
a transverse field.}

\end{figure}

For non-interacting systems, the problem has been addressed in many
works, including the pioneering works of Néel, Brown and Aharoni \citep{Neel_1949,Brown_PR1963,Aharoni_PhysRev1969}.
Here, our system consists of a chain of $\mathcal{N}$ ferromagnetic
nanoparticles, with a restriction to monodisperse assemblies of monodomain
MNP. The latter are represented by a single macrospin, $\bm{m}_{i}=m{\bf S}_{i}=n\mbox{\ensuremath{\mu}}_{B}{\bf S}_{i}$,
with a uniaxial anisotropy. All anisotropy axes are parallel and aligned
along the chain direction ${\bf e}_{z}$. We apply to the system an
external DC magnetic field ${\bf H}$ that can be either longitudinal
(along ${\bf e}_{z}$) or transverse (${\bf e}_{\perp}$). The two
situations are presented in Fig \ref{fig:1D-chain}.

The aim of this study is to investigate the variation of $T_{B}$
as a function of the interparticle distance $a$ and the applied external
dc field $H$. A naive approach would be to take an Arrhenius law
giving the relaxation time as 
\begin{equation}
\tau=\tau_{D}e^{\frac{\Delta\mathcal{E}}{k_{B}T}},\label{eq:relax_time}
\end{equation}
where $\Delta\mathcal{E}$ represents the energy barrier between the
two potential wells and $\tau_{D}$ the diffusion time. The maximum
blocking temperature is then defined with respect to the measuring
timescale through the expression: 
\begin{equation}
\tau_{{\rm mes}}=\tau_{D}e^{\frac{\Delta\mathcal{E}}{k_{B}T_{B}}}\ \Rightarrow T_{B}\propto\frac{\Delta\mathcal{E}}{k_{B}}.\label{eq:blocking_temp}
\end{equation}
We thus see that we first need to compute the variation of the energy
barrier $\Delta\mathcal{E}\left(a,H\right)$ in the presence of a
weak dipolar field. However, this simple approach suffers a major
drawback: by focusing only on the barrier, one neglects the dynamics
in the wells whilst Jönsson and Garc\'ia-Palacios showed that the
latter plays an essential role in the determination of the blocking
temperature \citep{Jonsson_Garcia-Palacios_PRB2001,Jonsson_Garcia_Palacios_EPL2001}.
In order to take this dynamics into account, we compute the relaxation
rate $\Gamma$ in the presence of weak dipolar interactions, in the
limit of intermediate to high damping using Langer's theory \citep{lan68prl,lan69ap}.
A similar approach was used by Braun for chains with exchange couplings
\citep{Braun_PRB1994} and is generalized here to take account of
the specificity of the long range nature of the dipolar interaction
\citep{Vernay_etal_acsucept_PRB2014}.

The paper is organized as follows: our model is introduced in Section
\ref{sec:Presentation-of-the} and we summarize the procedure to compute
the relaxation rate by following Langer's approach. The relaxation
rate in the longitudinal field case is presented in details in Section
\ref{sec:Longitudinal-field-case}, while the computation for the
transverse field case is done more briefly in Section \ref{sec:Transverse-field-case}.
Our results are disclosed in Section \ref{sec:Results}; we show in
particular the behavior of the relaxation rate as a function of the
field or the reduced energy barrier for different strengths of the
DDI. We also provide a detailed discussion of the behavior of the
blocking temperature in 1D MNP chains and interpret our results with
the help of a simple analytical formula. The paper closes with a conclusion
and outlook.

\section{Presentation of the model\label{sec:Presentation-of-the}}

The total energy of the chain is given by the addition of the anisotropy
energy, the Zeeman energy and the dipole-dipole interaction (DDI)
energy :
\begin{equation}
E=E_{{\rm anis}}+E_{{\rm Zeeman}}+E_{{\rm ddi}}\label{eq:energy}
\end{equation}
where
\[
E_{{\rm anis}}=-\sum_{i=1}^{\mathcal{N}}KV\left(S_{i}^{z}\right)^{2}
\]
\[
E_{{\rm Zeeman}}=-\sum_{i=1}^{\mathcal{N}}{\bf H}\cdot\bm{m}_{i}
\]

\begin{align*}
E_{{\rm ddi}} & =-\frac{\mu_{0}}{4\pi}\frac{m^{2}}{a^{3}}\sum_{i=1}^{\mathcal{N}}\sum_{i<j}{\bf S}_{i}\mathcal{\bm{D}}_{ij}{\bf S}_{j}
\end{align*}
with
\[
\mathcal{\bm{D}}_{ij}=\frac{3\cdot{\bf e}_{ij}\ {\bf e}_{ij}\cdot-1}{r_{ij}^{3}},\quad r_{ij}=\left\Vert {\bf r}_{i}-{\bf r}_{j}\right\Vert ,{\bf e}_{ij}=\frac{{\bf r}_{ij}}{r_{ij}}.
\]
The total energy can be measured in units of $2KV$ (anisotropy energy),
\emph{i.e.} $\mathcal{E}=E/\left(2KV\right)$. Accordingly, we define
the dimensionless physical parameters

\begin{equation}
h\equiv\frac{H}{H_{a}};\quad\xi\equiv\frac{\mu_{0}}{4\pi}\frac{m^{2}/a^{3}}{2KV}\label{eq:reduced_param}
\end{equation}
but keep the anisotropy parameter $k=1$ so as to be able to track
the anisotropy contribution in the subsequent developments. Consequently,
we write

\begin{equation}
\mathcal{E}=\sum_{i}^{\mathcal{N}}\mathcal{E}_{i}=-\frac{k}{2}\sum_{i=1}^{\mathcal{N}}S_{i,z}^{2}-\sum_{i=1}^{\mathcal{N}}\bm{h}\cdot{\bf S}_{i}-\xi\sum_{i=1}^{\mathcal{N}}\sum_{i<j}{\bf S}_{i}\mathcal{\bm{D}}_{ij}{\bf S}_{j}.\label{eq:total_energy}
\end{equation}
The estimation of the relaxation rate for an interacting chain of
nanoparticles may be relatively complicated to compute analytically
because of the discrete sums. However, in the case of a chain, the
lattice sum is straightforward to compute. We introduce the local
coordinates for each spin $\left(\theta_{i},\varphi_{i}\right)$,
such that $S_{i,z}\equiv\cos\theta_{i}$. We also define the compact
notation $V_{ij}\equiv V\left(\bm{r}_{i}-\bm{r}_{j}\right)=1/\left\Vert \bm{r}_{i}-\bm{r}_{j}\right\Vert ^{3}$
such that, owing to the symmetry with respect to rotations about the
chain's axis ($z$), we write the local energy as follows

\begin{equation}
\begin{array}{lll}
\mathcal{E}_{i} & = & -\bm{h}\cdot\bm{S}_{i}-\frac{k}{2}S_{i,z}^{2}\\
\\
 & - & \xi{\displaystyle \sum_{j\ne i}}V_{ij}\left[2S_{i,z}S_{j,z}-{\displaystyle \sum_{\alpha=x,y}}S_{i,\alpha}S_{j,\alpha}\right]
\end{array}\label{eq:EnergyLocal}
\end{equation}
and the total energy then reads

\begin{align}
\begin{array}{lll}
\mathcal{E} & = & -{\displaystyle \sum_{\alpha,i}}h^{\alpha}\;S_{i,\alpha}-\frac{k}{2}{\displaystyle \sum_{i}}S_{i,z}^{2}\\
\\
 & - & \xi{\displaystyle \sum_{i,j\ne i}}\left[2S_{i,z}S_{jz}-{\displaystyle \sum_{\alpha=x,y}}S_{i,\alpha}S_{j,\alpha}\right].
\end{array}\label{eq:EnergyTotal}
\end{align}
In the intermediate-to-high damping limit, the relaxation rate $\Gamma$
for an elementary process, \emph{i.e.} an escape from a metastable
state (m) to a state of lower energy, through a saddle point (s),
can be calculated with the help of Langer's theory \citep{lan68prl,lan69ap}.
$\Gamma$ can be put into the following compact form \citep{kac03epl}
\begin{equation}
\Gamma=\frac{\left|\kappa\right|}{2\pi}\frac{\mathcal{Z}_{s}}{\mathcal{Z}_{m}},\label{eq:Gamma_Langer}
\end{equation}
where $\kappa$ represents the attempt frequency to cross the barrier,
$\mathcal{Z}_{s}$ and $\mathcal{Z}_{m}$ are the partition functions
at, respectively, the saddle point (s) and the metastable state (m).
The explicit evaluation of Eq. (\ref{eq:Gamma_Langer}) thus requires
the analytical expression of the energy in the vicinity of the stable
state and the saddle point.

\section{Longitudinal case $\boldsymbol{h}=h\boldsymbol{e}_{z}$\label{sec:Longitudinal-field-case}}

\subsection{Energy barrier in the continuous limit in the longitudinal case}

In order to compute the energy barrier we need to determine the saddle
point. For this we compute the functional derivative $\delta\mathcal{E}_{i}/\delta S_{i,\alpha},\ \alpha=x,y,z$.
The two transverse components (\emph{i.e.} $\alpha=x,y$) yield the
constraints:
\[
\sum_{j}V_{ij}S_{j}^{\alpha=x,y}=0.
\]

This is consistent with the fact that the problem is symmetric with
respect to the $z$-axis. The longitudinal component $S_{z}$ contains
the most relevant information about the energy barrier:
\begin{equation}
S_{i,z}=-\frac{h}{k}-4\frac{\xi}{k}\sum_{j}V_{ij}S_{j,z}\label{eq:phi_z}
\end{equation}
which can be self-consistently solved leading to the following result
(to first order in $\xi$)
\begin{equation}
S_{i,z}^{\left(s\right)}=\cos\theta_{i}^{\left(s\right)}=-\frac{h}{k}\left[1-4\frac{\xi I_{i}}{k}\right],\label{eq:eta_saddle_longitu}
\end{equation}
with $I_{i}=\sum_{j}V_{ij}$ and where $\left(s\right)$ refers to
the saddle point. We have checked that for chains of more than 20
particles, the sum $I_{i}$ is nearly constant along the chain with
a maximum deviation at the edges of less than 5\%. This implies that
the lattice sum $I$ can be considered as independent of the site
at which it is computed. Its limit is given by the Riemann zeta function
$\zeta\left(3\right)\approx1.202$. Henceforth, in a first approximation
we consider chains that are sufficiently long to neglect edge effects,
such that $I_{i}=I$ and consequently $\theta_{i}^{\left(s\right)}=\theta_{s}$.
Substituting $S_{i,z}^{\left(s\right)}$ from Eq. (\ref{eq:eta_saddle_longitu})
back into Eq. (\ref{eq:EnergyLocal}) yields
\begin{equation}
\mathcal{E}_{s\parallel}^{\left(0\right)}\left(r\right)=\frac{h^{2}}{2k}\left(1-4\frac{\xi I}{k}\right).\label{eq:energy_saddle_longi}
\end{equation}
The energy at the (meta)stable state is gained by inserting $S_{i,z}^{\left(\pm\right)}=\pm1$
into Eq. (\ref{eq:EnergyLocal}), such that the energy barrier $\Delta\mathcal{E}_{\pm}$
with respect to the (meta)stable state, $\mathcal{E}_{\pm}^{\left(0\right)}=\pm h-\frac{k}{2}-2\xi I$
is given by
\begin{equation}
\Delta\mathcal{E}_{\pm}^{\parallel}=\frac{k}{2}\left(1\pm\frac{h}{k}\right)^{2}+2\xi I\left(1-\frac{h^{2}}{k^{2}}\right)\label{eq:energy_barrier}
\end{equation}
where the $\pm$ sign refers to the relative orientation of the field
with respect to the magnetic moments.

\subsection{Evaluation of the partition functions}

As is inherent to Langer's approach, the expression of the partition
functions in the vicinity of the saddle point and metastable state
are obtained by performing a second-order expansion of the energy.
For this purpose, it is easier to rewrite the equation of the energy
in Eq. (\ref{eq:EnergyLocal}) in spherical coordinates $(\theta,\varphi)$.
By definition of the extrema the first order derivative will not contribute
to the expansion once evaluated thereat.

The second-order expansion around the saddle point then reads
\[
\mathcal{E}_{s}^{\parallel}\simeq\mathcal{E}_{s\parallel}^{(0)}+\frac{1}{2}\left.\frac{\partial^{2}\mathcal{E}}{\partial\theta^{2}}\right|_{\theta_{s}}\left(\theta-\theta_{s}\right)^{2}.
\]
Inserting the value of $\cos\theta_{s}$ obtained in Eq. (\ref{eq:eta_saddle_longitu})
in the expression of the second derivative and keeping only the linear
terms in $\xi$, leads to
\begin{equation}
\mathcal{E}_{s}^{\parallel}\simeq\mathcal{E}_{s\parallel}^{\left(0\right)}+\frac{1}{2}\underbrace{k\left(\frac{h^{2}}{k^{2}}-1\right)\left[1+\frac{8\frac{h^{2}}{k^{2}}\frac{\xi I}{k}}{1-\frac{h^{2}}{k^{2}}}\right]}_{=-\lambda_{t}<0}\left(\theta-\theta_{s}\right)^{2}.\label{eq:energy_at_saddle_point}
\end{equation}
If the chain is sufficiently long we may neglect the edge effects
and assume that the deviation induced by the dipolar field is nearly
constant over the whole chain. Hence, the partition function at the
saddle point can be factorized and the partition function per spin
then reads
\[
\mathcal{Z}_{s,\parallel}=2\pi e^{-\beta\mathcal{E}_{s\parallel}^{\left(0\right)}}\sin\theta_{s}{\displaystyle \int}_{-\infty}^{+\infty}e^{-\beta\frac{\lambda_{t}}{2}\delta^{2}}d\delta
\]
which can be computed and then expanded in terms of $\xi$
\begin{equation}
\mathcal{Z}_{s,\parallel}=\left(2\pi\right)^{3/2}\sqrt{\frac{k_{B}T}{k}}\ e^{-\beta\mathcal{E}_{s\parallel}^{\left(0\right)}}.\label{eq:partition_fct_saddle}
\end{equation}

To first order, the dipolar field is only present because it shifts
the energy of the saddle point by hardening the anisotropy. In the
absence of $\xi$ in Eq. (\ref{eq:partition_fct_saddle}), one recovers
the standard expression for a single spin with a uniaxial anisotropy. 

Let us now compute $\mathcal{Z}_{-}$, the partition function near
the metastable state. Since there is a rotational invariance around
the $(Oz)$ axis, this part is easier to compute by using the Cartesian
coordinates
\[
\left\{ \begin{array}{ccc}
S_{z}^{2} & = & 1-S_{x}^{2}-S_{y}^{2}\\
\\
S_{z} & \simeq & -1+\frac{1}{2}S_{x}^{2}+\frac{1}{2}S_{y}^{2}
\end{array}\right.
\]
hence, the energy around the metastable point may be written as 
\[
\begin{array}{lll}
\mathcal{E}_{-} & \simeq & \underbrace{\left[-\frac{k}{2}+h-2\xi I\right]}_{=\mathcal{E}_{-}^{(0)}}\\
\\
 & + & \underbrace{k\frac{1-h/k}{2}\left[1+\frac{4\xi I/k}{1-h/k}\right]}_{=\mu}\left(S_{x}^{2}+S_{y}^{2}\right).
\end{array}
\]
Therefore the partition function is given by
\begin{equation}
\begin{array}{lll}
\mathcal{Z}_{-} & = & e^{-\beta\mathcal{E}_{-}^{\left(0\right)}}\left({\displaystyle \int_{-\infty}^{+\infty}}e^{-\beta\mu x^{2}}dx\right)^{2}\\
\\
 & = & e^{-\beta\mathcal{E}_{-}^{\left(0\right)}}\frac{2\pi k_{B}T}{k\left(1-\frac{h}{k}\right)\left[1+\frac{4\xi I/k}{1-h/k}\right]}.
\end{array}\label{eq:partition_fct_metastable}
\end{equation}

From Eqs. (\ref{eq:partition_fct_saddle}) and (\ref{eq:partition_fct_metastable}),
we obtain the ratio $\mathcal{Z}_{s,\parallel}/\mathcal{Z}_{-}$
\begin{equation}
\frac{\mathcal{Z}_{s,\parallel}}{\mathcal{Z}_{-}}=\sqrt{\frac{2\pi k}{k_{B}T}}\ e^{-\beta\Delta\mathcal{E}_{-}^{\parallel}}\ \left(1-\frac{h}{k}\right)\left[1+\frac{4\xi I/k}{1-\frac{h}{k}}\right]\label{eq:ratio_ZsZ1}
\end{equation}
where the energy barrier $\Delta\mathcal{E}_{-}$ is given by Eq.
(\ref{eq:energy_barrier}).

\subsection{Attempt frequency\label{subsec:Attempt-frequency}}

In order to complete the calculation of the relaxation rate of Eq.
(\ref{eq:Gamma_Langer}), we still have to compute the attempt frequency
$\kappa$. This is given by the first nonzero negative eigenvalue
of the transfer matrix. For computing it we proceed by writing the
Landau-Lifshitz equation in spherical coordinates
\begin{equation}
\left\{ \begin{array}{lcl}
\dot{\theta} & = & -\frac{1}{\sin\theta}\partial_{\varphi}\mathcal{E}-\alpha\partial_{\theta}\mathcal{E},\\
\\
\dot{\varphi} & = & -\frac{\alpha}{\sin\theta}\partial_{\varphi}\mathcal{E}+\partial_{\theta}\mathcal{E}
\end{array}\right.\label{eq:system_LL-1}
\end{equation}
where $\alpha$ is the damping parameter. We then make the expansion
of the coordinates $\left(\theta,\varphi\right)$ around the saddle
point $\left(\theta_{s},\varphi_{s}\right)$, \emph{i.e.} $\theta\simeq\theta_{s}+t,\varphi\simeq\varphi_{s}+p$
and next expand the energy to second order (call the result $\mathcal{E}_{s}^{\left(2\right)}$)
in $t,p$ upon which the Landau-Lifshitz equation becomes
\[
\left\{ \begin{array}{lcl}
\overset{\cdot}{t} & = & -\partial_{p}\mathcal{E}_{s}^{\left(2\right)}-\alpha\partial_{t}\mathcal{E}_{s}^{\left(2\right)},\\
\\
\dot{p} & = & -\alpha\partial_{p}\mathcal{E}_{s}^{\left(2\right)}+\partial_{t}\mathcal{E}_{s}^{\left(2\right)}.
\end{array}\right.
\]

The two equations above can be recast into the following matrix form
(using the notation $\eta_{i}=\left(t,p\right)$)
\[
\partial_{t}\eta_{i}=\sum_{j}M_{ij}\partial_{\eta_{j}}\mathcal{E}_{s}^{\left(2\right)}.
\]
Close to the saddle point, the energy may be expressed as $\mathcal{E}_{s}=\mathcal{E}_{s}^{\left(0\right)}+\frac{1}{2}\lambda_{t}t^{2}+\frac{1}{2}\lambda_{f}f^{2}$.
In the present case, we have $\lambda_{f}=0$ and $\lambda{}_{t}$
is defined in Eq. (\ref{eq:energy_at_saddle_point}). Hence, the eigenvalue
of the resulting matrix leads to 
\begin{equation}
\left|\kappa\right|=\alpha k\left(1-\frac{h^{2}}{k^{2}}\right)\left[1+\frac{8\frac{h^{2}}{k^{2}}\frac{\xi I}{k}}{1-\frac{h^{2}}{k^{2}}}\right].\label{eq:attempt_longi}
\end{equation}

\subsection{Relaxation rate in the longitudinal case}

The result in Eq. (\ref{eq:attempt_longi}) and the ratio in Eq. (\ref{eq:ratio_ZsZ1})
are used in Eq. (\ref{eq:Gamma_Langer}) to compute the relaxation
rate in longitudinal field, that is

\begin{equation}
\begin{array}{lll}
\Gamma_{-\to s} & = & \alpha k^{3/2}\sqrt{\frac{\beta}{2\pi}}\left(1-\frac{h^{2}}{k^{2}}\right)\left(1-\frac{h}{k}\right)\\
\\
 & \times & \left[1+\frac{8\frac{h^{2}}{k^{2}}\frac{\xi I}{k}}{1-\frac{h^{2}}{k^{2}}}\right]\left[1+\frac{4\xi I/k}{1-\frac{h}{k}}\right]e^{-\beta\Delta\mathcal{E}_{-}^{\parallel}}.
\end{array}\label{eq:gamma_1-s}
\end{equation}
By simply performing the change $h\to-h$, one can deduce the rate
$\Gamma_{+\to s}$
\begin{equation}
\begin{array}{lll}
\Gamma_{+\to s} & = & \alpha k^{3/2}\sqrt{\frac{\beta}{2\pi}}\left(1-\frac{h^{2}}{k^{2}}\right)\left(1+\frac{h}{k}\right)\\
\\
 & \times & \left[1+\frac{8\frac{h^{2}}{k^{2}}\frac{\xi I}{k}}{1-\frac{h^{2}}{k^{2}}}\right]\left[1+\frac{4\xi I/k}{1+\frac{h}{k}}\right]e^{-\beta\Delta\mathcal{E}_{+}^{\parallel}}.
\end{array}\label{eq:gamma_0-s}
\end{equation}
Adding up these two equations renders the total relaxation rate of
the chain's magnetic moment

\begin{equation}
\begin{array}{lll}
\Gamma_{\parallel} & = & \alpha k^{3/2}\sqrt{\frac{\beta}{2\pi}}\left(1-\frac{h^{2}}{k^{2}}\right)\\
\\
 & \times & \left\{ \left(1+\frac{h}{k}\right)\left[1+\frac{4\left(1-\frac{h}{k}+2\frac{h^{2}}{k^{2}}\right)\frac{\xi I}{k}}{1-\frac{h^{2}}{k^{2}}}\right]\ e^{-\beta\Delta\mathcal{E}_{+}^{\parallel}}\right.\\
\\
 & + & \left.\left(1-\frac{h}{k}\right)\left[1+\frac{4\left(1+\frac{h}{k}+2\frac{h^{2}}{k^{2}}\right)\frac{\xi I}{k}}{1-\frac{h^{2}}{k^{2}}}\right]\ e^{-\beta\Delta\mathcal{E}_{-}^{\parallel}}\right\} .
\end{array}\label{eq:overall_gamma_long}
\end{equation}

It can readily be checked that setting $\xi\to0$ in this expression
recovers the Néel-Brown result \citep{Aharoni_PhysRev1969}.

Eq. (\ref{eq:overall_gamma_long}) shows that the energy at the saddle
point changes due to the dipolar interaction as well as the external
magnetic field. The concomitant presence of the two contributions
leads to the additional cross term $\propto h^{2}\xi I$. In contrast,
even in the absence of the DC field, the energies of the two minima
are lowered by the same amount $2\xi I$ due to the dipolar interaction.
This variation can be absorbed in the definition of the anisotropy
constant $k$ by introducing the renormalized anisotropy constant
$k^{\prime}=k\left(1+\frac{4\xi I}{k}\right).$ This means that the
chain of interacting MNP would behave as a macrospin with effective
uniaxial anisotropy of constant $k^{\prime}$ with easy axis along
the chain.

\section{Transverse field \label{sec:Transverse-field-case}}

The external magnetic field is now normal to the chain axis and the
uniaxial anisotropy; we choose $\bm{h}=h\bm{e}_{x}$. The relaxation
rate for the transverse field is then obtained by following the same
procedure as described in detail in the previous section.

The energy in the continuum limit now reads
\begin{equation}
\begin{array}{lll}
\mathcal{E}_{i} & = & -hS_{i,x}-\frac{k}{2}S_{i,z}^{2}\\
\\
 & - & \xi{\displaystyle \sum_{j\ne i}}V_{ij}\left[2S_{i,z}S_{j,z}-{\displaystyle \sum_{\alpha=x,y}}S_{i,\alpha}S_{j,\alpha}\right],
\end{array}\label{eq:En-TF}
\end{equation}
and can be re-expressed in spherical coordinates as it is more convenient
for finding the extrema. The derivative with respect to $\varphi_{i}$
yields the following equation
\[
h\sin\varphi_{i}-2\xi\sum_{j\ne i}V_{ij}\sin\theta_{j}\sin\left(\varphi_{i}-\varphi_{j}\right)=0.
\]
Since the magnetic field is applied along the $x$ axis and the anisotropy
is along the $z$ axis, the effective field is necessarily in the
$xz$ plane and thereby we may simply set the azimuthal angle to zero,
\emph{i.e.} $\varphi_{i}=0$. Therefore, we derive the following simplified
equation for the polar angle $\theta_{i}$
\begin{align}
\begin{array}{lll}
0 & = & \cos\theta_{i}\left[-h+k\sin\theta_{i}\right]\\
\\
 & + & 2\xi{\displaystyle \sum_{j\ne i}}V_{ij}\left\{ 2\cos\theta_{j}\sin\theta_{i}+\sin\theta_{j}\cos\theta_{i}\right\} 
\end{array} & .\label{eq:1stDerivTheta-TF}
\end{align}

This equation may be solved perturbatively and to first order in $\xi$
it yields the position of the saddle point $\theta_{s}$ and the minimum
$\theta_{m}$ 
\begin{equation}
\left\{ \begin{array}{lll}
\theta_{s} & = & \frac{\pi}{2},\\
\\
\theta_{m} & = & \frac{h}{k}\left[1-\frac{6\xi I}{k}\right].
\end{array}\right.\label{eq:theta_min_sad_perp}
\end{equation}
Next, we can evaluate the energy at these two points leading to $\mathcal{E}_{s\perp}^{\left(0\right)}=-h+\xi I$
and $\mathcal{E}_{m}^{\left(0\right)}=-\frac{k}{2}-\frac{h^{2}}{2k}-2\xi I\left[1-\frac{3}{2}\frac{h^{2}}{k^{2}}\right]$,
and infer from the latter the energy barrier
\begin{equation}
\Delta\mathcal{E}^{\perp}=\mathcal{E}_{s\perp}^{\left(0\right)}-\mathcal{E}_{m}^{\left(0\right)}=\frac{k}{2}\left(1-\frac{h}{k}\right)^{2}+3\xi I\left(1-\frac{h^{2}}{k^{2}}\right).\label{eq:energy_barrier_perp}
\end{equation}
Since the addition of the external magnetic field now explicitly breaks
the rotational symmetry around the $z$-axis, the expansion of the
energy in the vicinity of the saddle point and the metastable state
contains a term in $\varphi$, 
\begin{equation}
\begin{array}{lll}
\mathcal{E}_{s}^{\perp} & \simeq & \mathcal{E}_{s\perp}^{\left(0\right)}+\frac{1}{2}\left[h-2\xi I\right]\left(\varphi-\varphi_{s}\right)^{2}\\
 & + & \frac{1}{2}\left[-k+h-2\xi I\right]\left(\theta-\theta_{s}\right)^{2},\\
\\
\mathcal{E}_{m} & \simeq & \mathcal{E}_{m}^{\left(0\right)}+\frac{1}{2}\frac{h^{2}}{k}\left[1-\frac{8\xi I}{k}\right]\left(\varphi-\varphi_{s}\right)^{2}\\
 & + & \frac{k}{2}\left[\left(1-\frac{h^{2}}{k^{2}}\right)+\frac{4\xi I}{k}\left(1+3\frac{h^{2}}{k^{2}}\right)\right]\left(\theta-\theta_{s}\right)^{2}.
\end{array}\label{eq:energy_at_saddle_point_perp}
\end{equation}

From this the partition function is obtained at the saddle point
\begin{equation}
\mathcal{Z}_{s,\perp}=\frac{2\pi k_{B}T}{k\sqrt{\frac{h}{k}\left(1-\frac{h}{k}\right)}}e^{-\beta\mathcal{E}_{s\perp}^{\left(0\right)}}\left(1+\frac{\xi I}{h}\frac{k-2h}{k-h}\right)\label{eq:partition_fct_sad_perp}
\end{equation}
and at the minimum (metastable state)
\begin{equation}
\mathcal{Z}_{m}=\frac{2\pi k_{B}T}{k\sqrt{1-\frac{h^{2}}{k^{2}}}}e^{-\beta\mathcal{E}_{m}^{\left(0\right)}}\left(1-\frac{4\xi I}{k}\frac{k^{2}+h^{2}}{k^{2}-h^{2}}\right).\label{eq:partition_fct_min_perp}
\end{equation}

Hence, to first order in the dipolar interaction ($\xi$), their ratio
reads
\begin{equation}
\begin{array}{lll}
\frac{\mathcal{Z}_{s,\perp}}{\mathcal{Z}_{m}} & = & \sqrt{\frac{1-\frac{h^{2}}{k^{2}}}{\frac{h}{k}\left(1-\frac{h}{k}\right)}}e^{-\beta\Delta\mathcal{E}^{\perp}}\\
\\
 & \times & \left[1+\xi I\frac{1+3\frac{h}{k}-2\left(\frac{h}{k}\right)^{2}+4\left(\frac{h}{k}\right)^{3}}{\frac{h}{k}\left(1-\frac{h^{2}}{k^{2}}\right)}\right].
\end{array}\label{eq:Zs_over_Zm_perp}
\end{equation}
Similarly to what was done in subsection \ref{subsec:Attempt-frequency},
the expression of the energy in the vicinity of the saddle point given
in Eq. (\ref{eq:energy_at_saddle_point_perp}), leads to the transition
matrix and the attempt frequency is obtained upon diagonalizing the
latter. This yields
\begin{equation}
\begin{array}{lll}
\kappa & = & \frac{k}{2}\left[\alpha\left(1-\frac{2h}{k}\right)+2\sqrt{\frac{h}{k}\left(1-\frac{h}{k}\right)+\frac{\alpha^{2}}{4}}\right]\\
\\
 & + & \xi I\left[2\alpha-\frac{1-\frac{2h}{k}}{\sqrt{\frac{h}{k}\left(1-\frac{h}{k}\right)+\frac{\alpha^{2}}{4}}}\right];
\end{array}\label{eq:attempt_freq_perp}
\end{equation}
which can symbolically written as $\kappa=\kappa^{\left(0\right)}+\xi I\kappa^{\left(1\right)}=\kappa^{\left(0\right)}\left(1+\xi I\frac{\kappa^{\left(1\right)}}{\kappa^{\left(0\right)}}\right)$,
where $\kappa^{\left(0\right)}$ is the result for the single spin
problem in a transverse field \citep{garetal99pre}. By collecting
the results of Eqs. (\ref{eq:Zs_over_Zm_perp}) and (\ref{eq:attempt_freq_perp})
and inserting them into Eq. (\ref{eq:Gamma_Langer}) we obtain the
expression of the relaxation rate in a transverse field
\begin{equation}
\begin{array}{lll}
\Gamma_{\perp} & = & \frac{\kappa^{\left(0\right)}}{2\pi}\sqrt{\frac{1-\frac{h^{2}}{k^{2}}}{\frac{h}{k}\left(1-\frac{h}{k}\right)}}e^{-\beta\Delta\mathcal{E}^{\perp}}\\
\\
 & \times & \left[1+\xi I\left(\frac{1+3\frac{h}{k}-2\left(\frac{h}{k}\right)^{2}+4\left(\frac{h}{k}\right)^{3}}{\frac{h}{k}\left(1-\frac{h^{2}}{k^{2}}\right)}+\frac{\kappa^{\left(1\right)}}{\kappa^{\left(0\right)}}\right)\right].
\end{array}\label{eq:overall_gamma_perp}
\end{equation}

It can be readily checked that upon setting $\xi=0$, one recovers
the relaxation rate of an isolated spin \citep{garetal99pre}.

\section{Discussion of the results\label{sec:Results}}

Let us now consider a chain of monodisperse iron-cobalt particles
with a radius $R=4\ {\rm nm}$, an effective uniaxial anisotropy $K=4.5\times10^{4}\ {\rm J/m^{3}}$,
and $M_{s}=1.162\times10^{6}{\rm A/m}$. The reduced anisotropy barrier
with respect to thermal energy is given by
\begin{equation}
\sigma=\frac{KV}{k_{B}T},\label{eq:sigma_anisotropy}
\end{equation}
and is relatively large $\sigma\left(300{\rm K}\right)\simeq3$ at
room temperature. The free diffusion time $\tau_{D}$, defined by
\begin{equation}
\tau_{D}=\frac{M_{s}}{2\alpha\gamma_{g}K},\label{eq:free_diffusion}
\end{equation}
where $\gamma_{g}$ is the gyromagnetic ratio. With these specific
parameters and for intermediate-to-high damping (namely $1\lesssim\alpha\lesssim10$)
$\tau_{D}$ varies between $7\times10^{-11}{\rm s}$ and $7\times10^{-12}{\rm s}$.
As already mentioned earlier, the chain is assumed to be long enough
so as to neglect the edge effects. We have checked that this assumption
becomes valid when the chain consists of more than 20 nanoparticles.

\subsection{Relaxation rate $\Gamma$}

\begin{figure}
\begin{centering}
(a)\includegraphics[width=0.95\columnwidth]{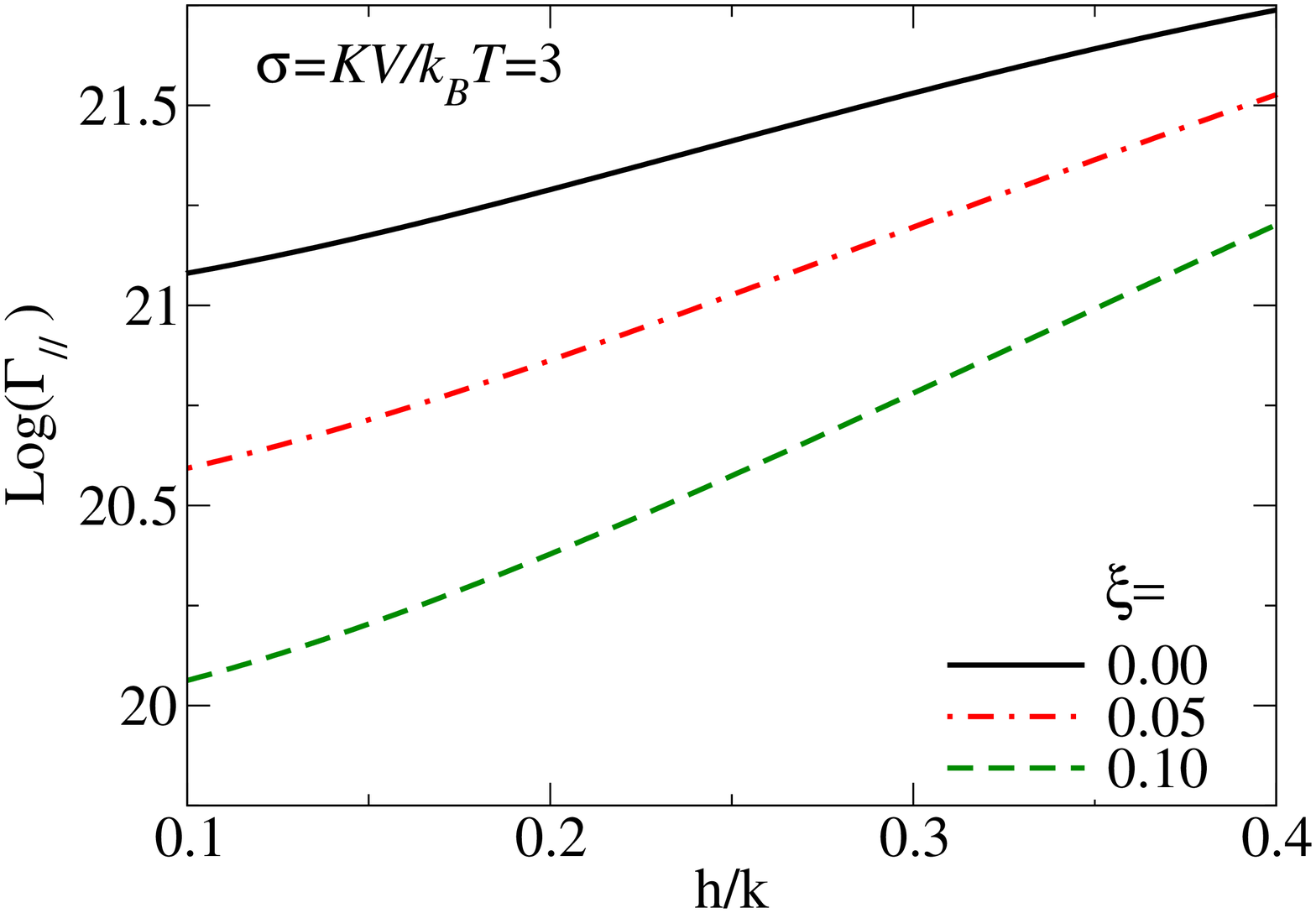}
\par\end{centering}
\begin{centering}
(b)\includegraphics[width=0.95\columnwidth]{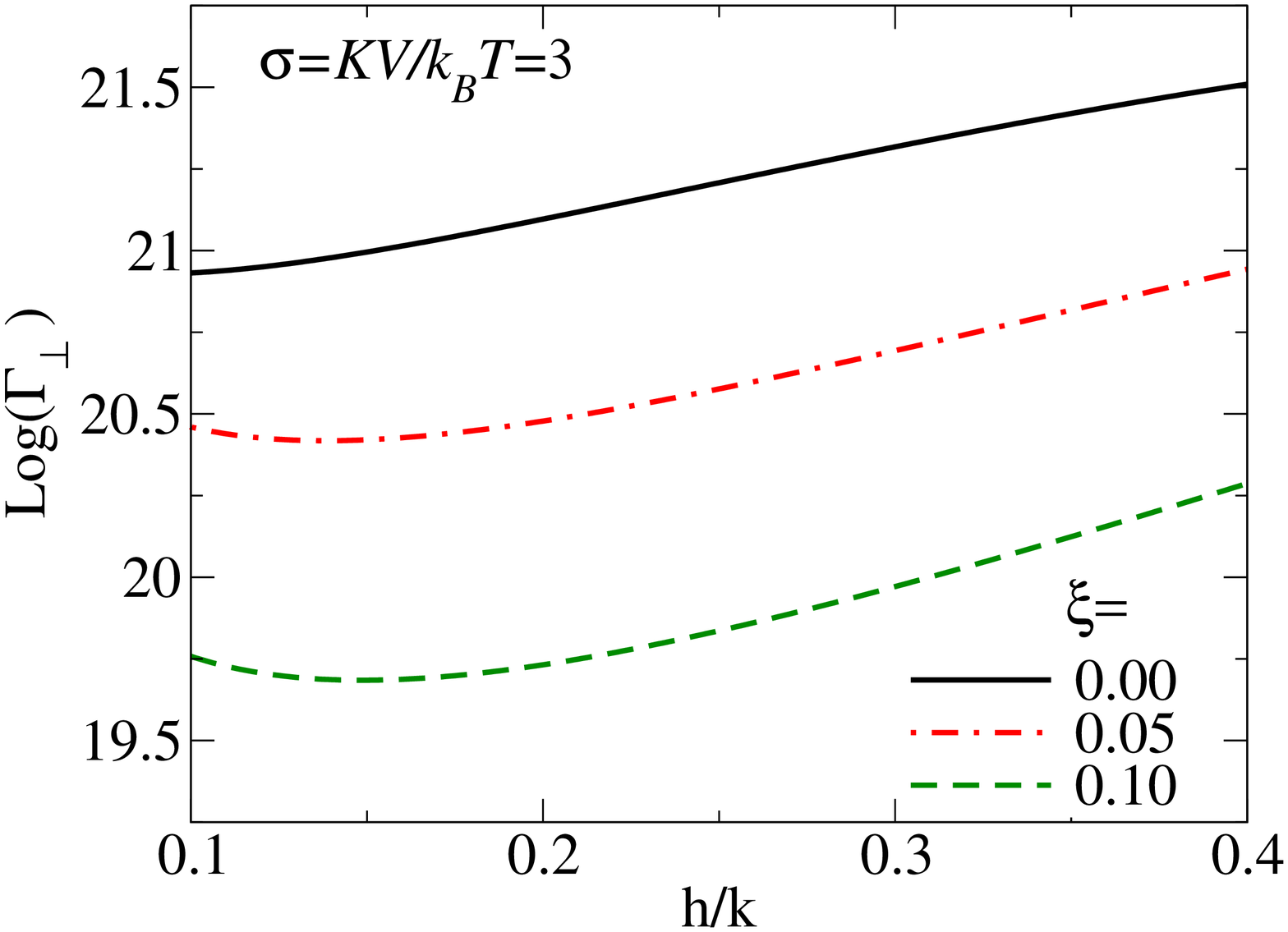}
\par\end{centering}
\caption{Relaxation rate as a function of the external dc field for different
values of the dipolar interaction $\xi$. The damping parameter is
$\alpha=1$ for both sets.\label{fig:Relaxation-rate-field}}
\end{figure}

In the expressions of the longitudinal and transverse relaxation rates
given in Eqs. (\ref{eq:overall_gamma_long}) and (\ref{eq:overall_gamma_perp}),
using the respective energy barriers in (\ref{eq:energy_barrier})
and (\ref{eq:energy_barrier_perp}), we see that the arguments of
the exponential functions are primarily governed by the zero field
energy barrier $\sigma=KV/k_{B}T$. Besides, if we inspect more closely
these two arguments for $h\to0$, we realize that taking the interparticle
dipolar interaction into account is equivalent to doing a renormalization
of the anisotropy. For a 1D chain, the dipolar field can indeed only
be along the chain and thus it merely brings an additional rigidity
to the magnetic system. This implies that the relaxation rate decreases
as the dipolar interaction increases. This can be checked in Figs.
\ref{fig:Relaxation-rate-field}-(a) and (b) where the logarithm of
the relaxation rate is plotted against the field for different dipolar
strengths $\xi$.

The relaxation rate $\Gamma_{\parallel}$ is mostly given by $\exp\left(-\beta\Delta\mathcal{E}_{-}^{\parallel}\right)$
and its prefactor as $h$ increases. In the prefactor, $\xi$ is coupled
to $h$ only via positive terms. This implies a monotonic behavior
of $\log\left(\Gamma_{\parallel}\right)$ as a function of $h$. 

In contrast, the prefactor of the relaxation rate $\Gamma_{\perp}$
in transverse field has a more involved expression which is a non
monotonic function of $h$. As a consequence, and as it can be seen
in Fig. \ref{fig:Relaxation-rate-field}-(b), we observe a competition
between the external and the dipolar fields. For finite $\xi$, the
relaxation rate first decreases at low fields and then increases when
$h$ overcomes the additional rigidity brought in by $\xi$.

\begin{figure}
\begin{centering}
(a)\includegraphics[width=0.95\columnwidth]{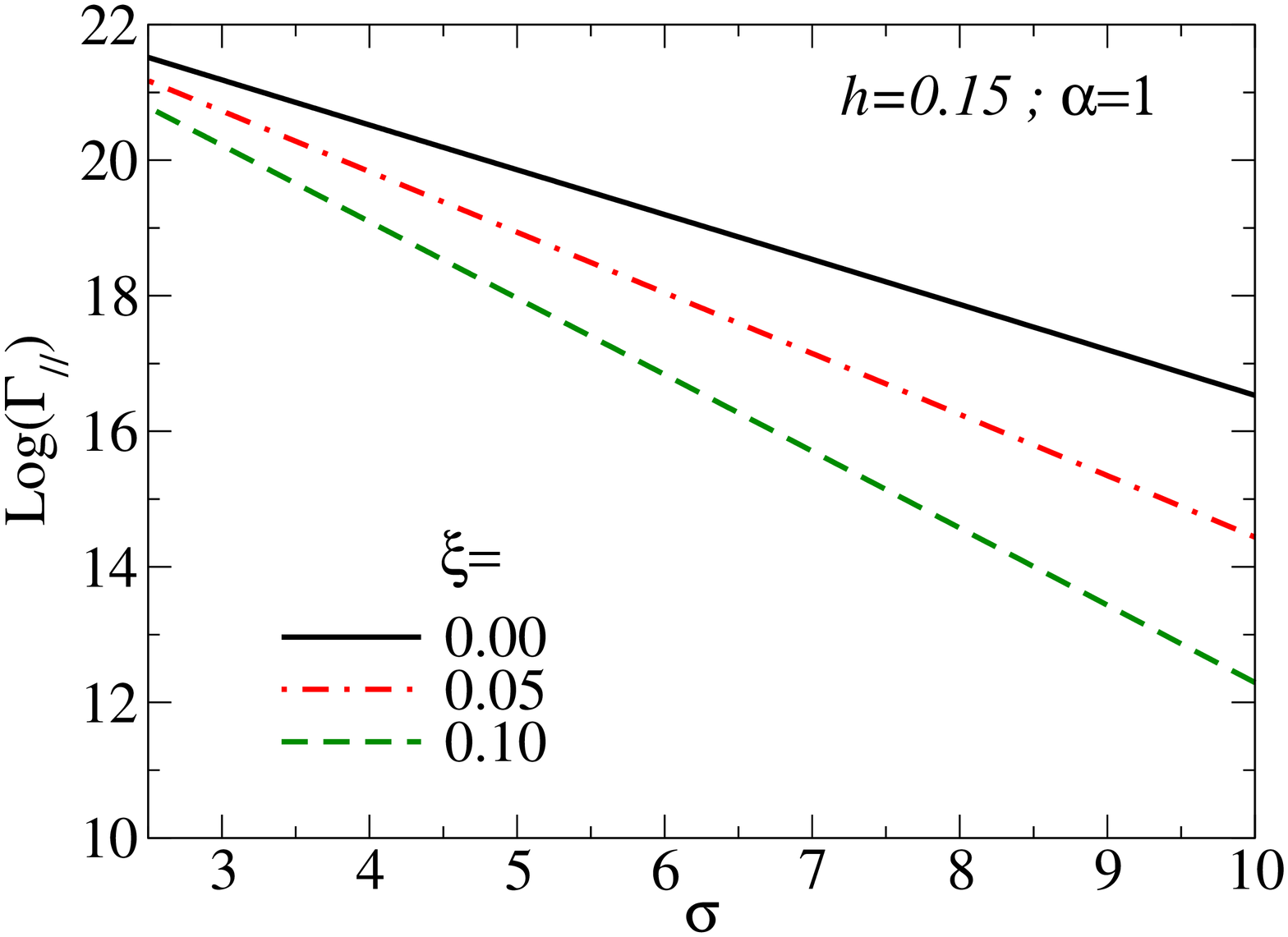}
\par\end{centering}
\begin{centering}
(b)\includegraphics[width=0.95\columnwidth]{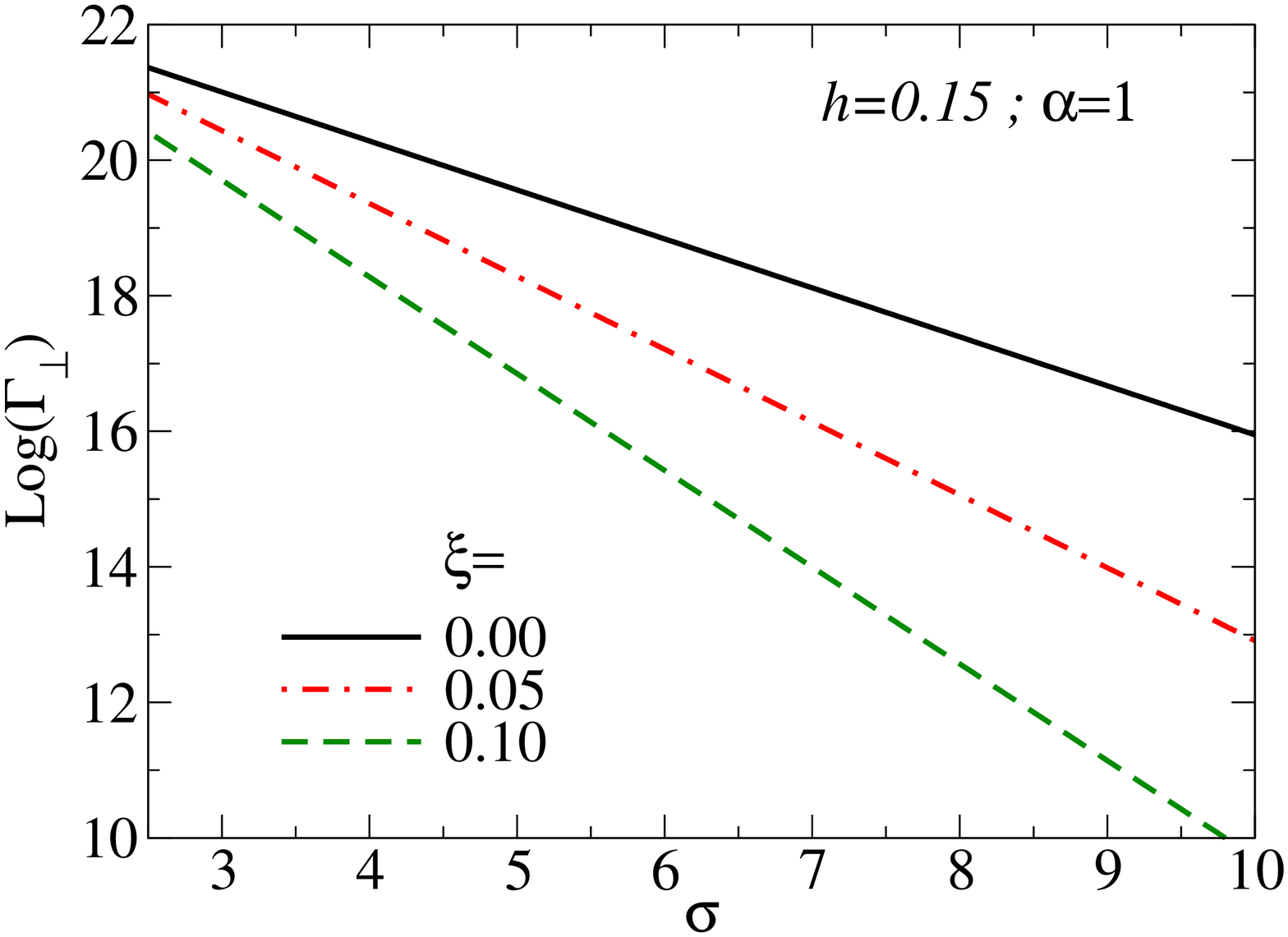}
\par\end{centering}
\caption{Relaxation rate as a function of $\sigma=KV/k_{B}T$ for different
values of the dipolar interaction $\xi$ and for a finite external
field $\frac{h}{k}=0.15$.\label{fig:Relaxation-rate-sigma}}
\end{figure}

When the anisotropy barrier $\sigma$ becomes large enough, say $\sigma\gtrsim2.5$,
the behavior of $\log\left(\Gamma_{\parallel}\right)$ and $\log\left(\Gamma_{\perp}\right)$
as a function of $\sigma$ is nearly linear as seen in Figs. \ref{fig:Relaxation-rate-sigma}--(a)
and (b). The effect of the dipolar field is mainly to increase the
(anisotropy) rigidity as explained earlier and is more pronounced
in the transverse field case. This can be understood by inspecting
the expressions of the energy barriers given by Eqs. (\ref{eq:energy_barrier})
and (\ref{eq:energy_barrier_perp}): the numerical prefactor in front
of $\xi$ is larger in $\Delta\mathcal{E}_{\perp}$. 

\subsection{Blocking temperature $T_{B}$}

The blocking temperature is obtained by solving Eq. (\ref{eq:blocking_temp})
for an experiment specific time $\tau_{{\rm mes}}$. In the case of
magnetic nanoparticles, the characterization of magnetic properties
can be achieved through SQUID experiments with a typical measuring
time $\tau_{{\rm SQUID}}=10^{-2}{\rm s}$.

\begin{figure}
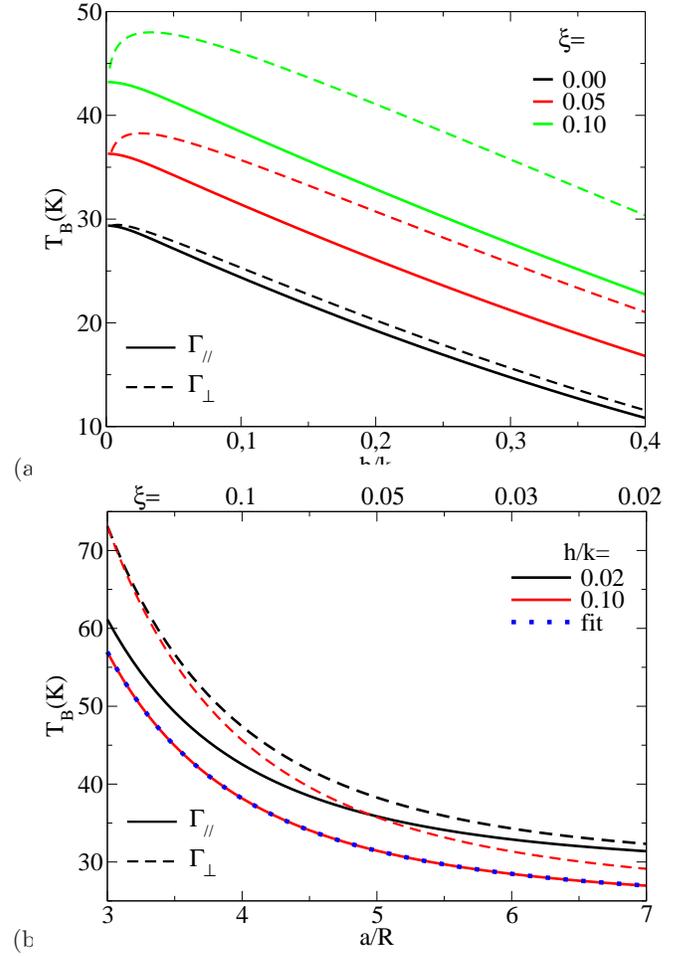

\begin{centering}
(a)\includegraphics[width=0.95\columnwidth]{fig4a}
\par\end{centering}
\begin{centering}
(b)\includegraphics[width=0.95\columnwidth]{fig4b}
\par\end{centering}
\caption{(a) $T_{B}$ as a function of the reduced external field $h$ for
different values of $\xi$. (b) $T_{B}$ as a function of the interparticle
distance $a$ (in units of the particle radius $R$) for different
values of the applied field $h$. For both panels $\alpha=1$, the
dashed lines represent the transverse field case, the continuous line
shows the longitudinal field case. The measuring time is typical of
SQUID experiment $\tau_{{\rm SQUID}}=10^{-2}{\rm s}$.\label{fig:Tb_of_h_and_a}}
\end{figure}

The blocking temperature $T_{B}$ is plotted in Fig. \ref{fig:Tb_of_h_and_a}-(a)
as a function of the external field for various dipolar strengths.
When $h\to0$, one sees that $T_{B}^{\perp}$ and $T_{B}^{\parallel}$
nearly coincide as expected. In the longitudinal case, the blocking
temperature is a decreasing function of $h$. Indeed, in this case,
as already stated earlier, the prefactor of the relaxation rate does
not play much of a role as it is a monotonic function of $h$ (for
low field and low $\xi$). Hence, the addition of the external field
for a finite $\xi$ is seen in the energy barrier and facilitates
the magnetization reversal by making more pronounced saddle points.
All in all, this results in a reduced $T_{B}$ as $h$ increases.

The situation in transverse field is more subtle: it involves both
the energy barrier and the prefactor of the relaxation rate. At a
relatively high field, the physics is governed by the energy barrier
which is strongly reduced, and hence $T_{B}$ decreases by increasing
$h$. In the low-field regime $h/k<0.05$, $h$ couples to $\xi$
and the competition that occurs between the dipolar (longitudinal)
field and the external (transverse) field leads to a nonmonotonic
behavior of $T_{B}$: as $h$ increases, $T_{B}^{\perp}$ first increases,
and then decreases when a critical value of the transverse field is
reached.

Fig. \ref{fig:Tb_of_h_and_a}-(b) shows the blocking temperature against
the interparticle distance $a$ for two different values of the external
field. The interparticle distance range has been chosen so that we
remain in the limit $\xi\ll1$, indeed even for $a/R=3$, we have
$\xi\approx0.232$. The behavior of $T_{B}^{\perp}$ and $T_{B}^{\parallel}$
is in line with the previous observations. Indeed, knowing that $\xi\propto a^{-3}$,
we see that if $a$ is increased the effective magnetic anisotropy
of the chain is reduced and the blocking temperature is lowered since
the energy barrier can be more easily overcome by the magnetic moments.

Indeed, for both Figs \ref{fig:Tb_of_h_and_a}-(a) and (b) the overall
behavior of $T_{B}$ can be understood if one simply uses the Arrhenius
law since then Eq. (\ref{eq:blocking_temp}) can be easily solved
to obtain (to the leading order in $h/k$)
\begin{equation}
T_{B}=\frac{1}{\log\left(\frac{\tau_{{\rm mes}}}{\tau_{D}}\right)}\left\{ \begin{array}{lll}
\frac{k}{2}\left(1-2\frac{h}{k}\right)+2\xi I(\bm{r}) & , & \ \parallel\\
\\
\frac{k}{2}\left(1-2\frac{h}{k}\right)+3\xi I(\bm{r}) & , & \ \perp
\end{array}\right.\label{eq:arrhenius_fit}
\end{equation}
Eq. (\ref{eq:arrhenius_fit}) correctly accounts for the behavior
of $T_{B}\left(a\right)$. For example, in Fig. \ref{fig:Tb_of_h_and_a}-(b)
we plot in blue points the result of fitting one of the curves using
Eq. (\ref{eq:arrhenius_fit}). This yields an excellent agreement
with $T_{B}=T_{B}^{\infty}+\frac{C}{(a/R)^{3}}$, and $T_{B}^{\infty}=24.4{\rm K}$,
$C=879.5{\rm K}$.

Concerning the interpretation of Fig. \ref{fig:Tb_of_h_and_a}-(a),
we see by inspecting Eq. (\ref{eq:arrhenius_fit}) that, if one neglects
the effect of the dynamics within the wells, one should expect a linearly
decreasing behavior for $T_{B}\left(h\right)$ with the same coefficient
for both $T_{B}^{\perp}$ and $T_{B}^{\parallel}$. This corresponds
exactly to what is shown in Fig. \ref{fig:Tb_of_h_and_a}-(a).

\section{conclusion and outlook}

On the basis of our analytical developments, we have shown in the
present work that for one-dimensional chains of nanoparticles, the
dipolar interaction mainly acts as an additional effective uniaxial
anisotropy. In essence, this renormalized rigidity implies an increase
of the relaxation rate. To be more specific, two situations can be
further analyzed: i) at relatively high fields and high barrier the
physics is largely dominated by the energy barrier physics and thus
by the argument of the exponential; ii) at lower fields, we have observed
a subtle role of the prefactor that can, for instance, lead to a non-monotonic
behavior of $\log\Gamma_{\perp}$. This is a prototypical example
that highlights the fact that the magnetization dynamics within the
well cannot be neglected and simply analyze the physics using the
Arrhenius law. This issue is particularly important in the context
of realistic experimental situations where one as to investigate the
dynamics of samples with (weakly) interacting chains. In these cases,
the inter-chain coupling can indeed be viewed as an effective field
with a transverse component \citep{ToulemonEtal_AFM2016} that will
affect the dynamics and can lead to non-monotonic behavior of the
relaxation rate.

\bibliography{biblio_FV}

\begin{thebibliography}{20}
\expandafter\ifx\csname natexlab\endcsname\relax\def\natexlab#1{#1}\fi
\expandafter\ifx\csname bibnamefont\endcsname\relax
  \def\bibnamefont#1{#1}\fi
\expandafter\ifx\csname bibfnamefont\endcsname\relax
  \def\bibfnamefont#1{#1}\fi
\expandafter\ifx\csname citenamefont\endcsname\relax
  \def\citenamefont#1{#1}\fi
\expandafter\ifx\csname url\endcsname\relax
  \def\url#1{\texttt{#1}}\fi
\expandafter\ifx\csname urlprefix\endcsname\relax\def\urlprefix{URL }\fi
\providecommand{\bibinfo}[2]{#2}
\providecommand{\eprint}[2][]{\url{#2}}

\bibitem[{\citenamefont{{Z. Sabsabi, F. Vernay, O. Iglesias, H.
  Kachkachi}}(2013)}]{sabsabietal13prb}
\bibinfo{author}{\bibnamefont{{Z. Sabsabi, F. Vernay, O. Iglesias, H.
  Kachkachi}}}, \bibinfo{journal}{Phys. Rev. B} \textbf{\bibinfo{volume}{88}},
  \bibinfo{pages}{104424} (\bibinfo{year}{2013}),
  \urlprefix\url{http://link.aps.org/doi/10.1103/PhysRevB.88.104424}.

\bibitem[{\citenamefont{Vernay et~al.}(2014)\citenamefont{Vernay, Sabsabi, and
  Kachkachi}}]{Vernay_etal_acsucept_PRB2014}
\bibinfo{author}{\bibfnamefont{F.}~\bibnamefont{Vernay}},
  \bibinfo{author}{\bibfnamefont{Z.}~\bibnamefont{Sabsabi}}, \bibnamefont{and}
  \bibinfo{author}{\bibfnamefont{H.}~\bibnamefont{Kachkachi}},
  \bibinfo{journal}{Phys. Rev. B} \textbf{\bibinfo{volume}{90}},
  \bibinfo{pages}{094416} (\bibinfo{year}{2014}),
  \urlprefix\url{http://link.aps.org/doi/10.1103/PhysRevB.90.094416}.

\bibitem[{\citenamefont{Toulemon et~al.}(2011)\citenamefont{Toulemon, Pichon,
  Catto{\"e}n, Man, and B{\'e}gin-Colin}}]{Toulemon_etal_2d_2011}
\bibinfo{author}{\bibfnamefont{D.}~\bibnamefont{Toulemon}},
  \bibinfo{author}{\bibfnamefont{B.~P.} \bibnamefont{Pichon}},
  \bibinfo{author}{\bibfnamefont{X.}~\bibnamefont{Catto{\"e}n}},
  \bibinfo{author}{\bibfnamefont{M.~W.~C.} \bibnamefont{Man}},
  \bibnamefont{and}
  \bibinfo{author}{\bibfnamefont{S.}~\bibnamefont{B{\'e}gin-Colin}},
  \bibinfo{journal}{Chem. Commun.} \textbf{\bibinfo{volume}{47}},
  \bibinfo{pages}{11954} (\bibinfo{year}{2011}),
  \urlprefix\url{http://dx.doi.org/10.1039/C1CC14661K}.

\bibitem[{\citenamefont{Pauly et~al.}(2011)\citenamefont{Pauly, Pichon, Albouy,
  Fleutot, Leuvrey, Trassin, Gallani, and
  Begin-Colin}}]{Pauly_Pichon_etal_2dassembly_JMatChem_2011}
\bibinfo{author}{\bibfnamefont{M.}~\bibnamefont{Pauly}},
  \bibinfo{author}{\bibfnamefont{B.~P.} \bibnamefont{Pichon}},
  \bibinfo{author}{\bibfnamefont{P.-A.} \bibnamefont{Albouy}},
  \bibinfo{author}{\bibfnamefont{S.}~\bibnamefont{Fleutot}},
  \bibinfo{author}{\bibfnamefont{C.}~\bibnamefont{Leuvrey}},
  \bibinfo{author}{\bibfnamefont{M.}~\bibnamefont{Trassin}},
  \bibinfo{author}{\bibfnamefont{J.-L.} \bibnamefont{Gallani}},
  \bibnamefont{and}
  \bibinfo{author}{\bibfnamefont{S.}~\bibnamefont{Begin-Colin}},
  \bibinfo{journal}{J. Mater. Chem.} \textbf{\bibinfo{volume}{21}},
  \bibinfo{pages}{16018} (\bibinfo{year}{2011}),
  \urlprefix\url{http://dx.doi.org/10.1039/C1JM12012C}.

\bibitem[{\citenamefont{Pauly et~al.}(2012)\citenamefont{Pauly, Pichon,
  Panissod, Fleutot, Rodriguez, Drillon, and
  Begin-Colin}}]{Pauly_Pichon_etal_2dassembly_JMatChem_2012}
\bibinfo{author}{\bibfnamefont{M.}~\bibnamefont{Pauly}},
  \bibinfo{author}{\bibfnamefont{B.~P.} \bibnamefont{Pichon}},
  \bibinfo{author}{\bibfnamefont{P.}~\bibnamefont{Panissod}},
  \bibinfo{author}{\bibfnamefont{S.}~\bibnamefont{Fleutot}},
  \bibinfo{author}{\bibfnamefont{P.}~\bibnamefont{Rodriguez}},
  \bibinfo{author}{\bibfnamefont{M.}~\bibnamefont{Drillon}}, \bibnamefont{and}
  \bibinfo{author}{\bibfnamefont{S.}~\bibnamefont{Begin-Colin}},
  \bibinfo{journal}{J. Mater. Chem.} \textbf{\bibinfo{volume}{22}},
  \bibinfo{pages}{6343} (\bibinfo{year}{2012}),
  \urlprefix\url{http://dx.doi.org/10.1039/C2JM15797G}.

\bibitem[{\citenamefont{Mann}(1985)}]{mann1985structure}
\bibinfo{author}{\bibfnamefont{S.}~\bibnamefont{Mann}}, in
  \emph{\bibinfo{booktitle}{Magnetite biomineralization and magnetoreception in
  organisms}} (\bibinfo{publisher}{Springer}, \bibinfo{year}{1985}), pp.
  \bibinfo{pages}{311--332}.

\bibitem[{\citenamefont{Bazylinski and
  Frankel}(2004)}]{bazylinski2004magnetosome}
\bibinfo{author}{\bibfnamefont{D.~A.} \bibnamefont{Bazylinski}}
  \bibnamefont{and} \bibinfo{author}{\bibfnamefont{R.~B.}
  \bibnamefont{Frankel}}, \bibinfo{journal}{Nature Reviews Microbiology}
  \textbf{\bibinfo{volume}{2}}, \bibinfo{pages}{217} (\bibinfo{year}{2004}),
  \urlprefix\url{https://doi.org/10.1038/nrmicro842}.

\bibitem[{\citenamefont{Charilaou et~al.}(2011)\citenamefont{Charilaou,
  Winklhofer, and Gehring}}]{Charilaou_FMR_chains_JAP2011}
\bibinfo{author}{\bibfnamefont{M.}~\bibnamefont{Charilaou}},
  \bibinfo{author}{\bibfnamefont{M.}~\bibnamefont{Winklhofer}},
  \bibnamefont{and} \bibinfo{author}{\bibfnamefont{A.~U.}
  \bibnamefont{Gehring}}, \bibinfo{journal}{Journal of Applied Physics}
  \textbf{\bibinfo{volume}{109}}, \bibinfo{pages}{093903}
  (\bibinfo{year}{2011}), \eprint{https://doi.org/10.1063/1.3581103},
  \urlprefix\url{https://doi.org/10.1063/1.3581103}.

\bibitem[{\citenamefont{Myrovali et~al.}(2016)\citenamefont{Myrovali, Maniotis,
  Makridis, Terzopoulou, Ntomprougkidis, Simeonidis, Sakellari, Kalogirou,
  Samaras, Salikhov et~al.}}]{myrovali2016arrangement}
\bibinfo{author}{\bibfnamefont{E.}~\bibnamefont{Myrovali}},
  \bibinfo{author}{\bibfnamefont{N.}~\bibnamefont{Maniotis}},
  \bibinfo{author}{\bibfnamefont{A.}~\bibnamefont{Makridis}},
  \bibinfo{author}{\bibfnamefont{A.}~\bibnamefont{Terzopoulou}},
  \bibinfo{author}{\bibfnamefont{V.}~\bibnamefont{Ntomprougkidis}},
  \bibinfo{author}{\bibfnamefont{K.}~\bibnamefont{Simeonidis}},
  \bibinfo{author}{\bibfnamefont{D.}~\bibnamefont{Sakellari}},
  \bibinfo{author}{\bibfnamefont{O.}~\bibnamefont{Kalogirou}},
  \bibinfo{author}{\bibfnamefont{T.}~\bibnamefont{Samaras}},
  \bibinfo{author}{\bibfnamefont{R.}~\bibnamefont{Salikhov}},
  \bibnamefont{et~al.}, \bibinfo{journal}{Scientific reports}
  \textbf{\bibinfo{volume}{6}}, \bibinfo{pages}{37934} (\bibinfo{year}{2016}),
  \urlprefix\url{https://doi.org/10.1038/srep37934}.

\bibitem[{\citenamefont{N{\'e}el}(1950)}]{Neel_1949}
\bibinfo{author}{\bibfnamefont{L.}~\bibnamefont{N{\'e}el}},
  \bibinfo{journal}{{J. Phys. Radium}} \textbf{\bibinfo{volume}{11}},
  \bibinfo{pages}{49} (\bibinfo{year}{1950}),
  \urlprefix\url{https://hal.archives-ouvertes.fr/jpa-00234217}.

\bibitem[{\citenamefont{Brown}(1963)}]{Brown_PR1963}
\bibinfo{author}{\bibfnamefont{W.~F.} \bibnamefont{Brown}},
  \bibinfo{journal}{Phys. Rev.} \textbf{\bibinfo{volume}{130}},
  \bibinfo{pages}{1677} (\bibinfo{year}{1963}),
  \urlprefix\url{https://link.aps.org/doi/10.1103/PhysRev.130.1677}.

\bibitem[{\citenamefont{Aharoni}(1969)}]{Aharoni_PhysRev1969}
\bibinfo{author}{\bibfnamefont{A.}~\bibnamefont{Aharoni}},
  \bibinfo{journal}{Phys. Rev.} \textbf{\bibinfo{volume}{177}},
  \bibinfo{pages}{793} (\bibinfo{year}{1969}),
  \urlprefix\url{http://link.aps.org/doi/10.1103/PhysRev.177.793}.

\bibitem[{\citenamefont{J\"onsson and
  Garc\'{\i}a-Palacios}(2001)}]{Jonsson_Garcia-Palacios_PRB2001}
\bibinfo{author}{\bibfnamefont{P.~E.} \bibnamefont{J\"onsson}}
  \bibnamefont{and} \bibinfo{author}{\bibfnamefont{J.~L.}
  \bibnamefont{Garc\'{\i}a-Palacios}}, \bibinfo{journal}{Phys. Rev. B}
  \textbf{\bibinfo{volume}{64}}, \bibinfo{pages}{174416}
  (\bibinfo{year}{2001}),
  \urlprefix\url{https://link.aps.org/doi/10.1103/PhysRevB.64.174416}.

\bibitem[{\citenamefont{J\"onsson and
  Garc{\'{\i}}a-Palacios}(2001)}]{Jonsson_Garcia_Palacios_EPL2001}
\bibinfo{author}{\bibfnamefont{P.~E.} \bibnamefont{J\"onsson}}
  \bibnamefont{and} \bibinfo{author}{\bibfnamefont{J.~L.}
  \bibnamefont{Garc{\'{\i}}a-Palacios}}, \bibinfo{journal}{Europhysics Letters
  ({EPL})} \textbf{\bibinfo{volume}{55}}, \bibinfo{pages}{418}
  (\bibinfo{year}{2001}),
  \urlprefix\url{https://doi.org/10.1209%2Fepl%2Fi2001-00430-0}.

\bibitem[{\citenamefont{Langer}(1968)}]{lan68prl}
\bibinfo{author}{\bibfnamefont{J.~S.} \bibnamefont{Langer}},
  \bibinfo{journal}{Phys. Rev. Lett.} \textbf{\bibinfo{volume}{21}},
  \bibinfo{pages}{973} (\bibinfo{year}{1968}),
  \urlprefix\url{http://link.aps.org/doi/10.1103/PhysRevLett.21.973}.

\bibitem[{\citenamefont{{J.S. Langer, \textit{Statistical theory of the decay
  of metastable states}}}(1969)}]{lan69ap}
\bibinfo{author}{\bibnamefont{{J.S. Langer, \textit{Statistical theory of the
  decay of metastable states}}}}, \bibinfo{journal}{{Ann. Phys. (N.Y.)}}
  \textbf{\bibinfo{volume}{54}}, \bibinfo{pages}{258} (\bibinfo{year}{1969}).

\bibitem[{\citenamefont{Braun}(1994)}]{Braun_PRB1994}
\bibinfo{author}{\bibfnamefont{H.-B.} \bibnamefont{Braun}},
  \bibinfo{journal}{Phys. Rev. B} \textbf{\bibinfo{volume}{50}},
  \bibinfo{pages}{16501} (\bibinfo{year}{1994}),
  \urlprefix\url{https://link.aps.org/doi/10.1103/PhysRevB.50.16501}.

\bibitem[{\citenamefont{{Kachkachi, H.}}(2003)}]{kac03epl}
\bibinfo{author}{\bibnamefont{{Kachkachi, H.}}}, \bibinfo{journal}{Europhys.
  Lett.} \textbf{\bibinfo{volume}{62}}, \bibinfo{pages}{650}
  (\bibinfo{year}{2003}),
  \urlprefix\url{https://doi.org/10.1209/epl/i2003-00423-y}.

\bibitem[{\citenamefont{{D.A. Garanin, E.C. Kennedy, D.S.F. Crothers, and W.T.
  Coffey}}(1999)}]{garetal99pre}
\bibinfo{author}{\bibnamefont{{D.A. Garanin, E.C. Kennedy, D.S.F. Crothers, and
  W.T. Coffey}}}, \bibinfo{journal}{{Phys. Rev. E}}
  \textbf{\bibinfo{volume}{60}}, \bibinfo{pages}{6499} (\bibinfo{year}{1999}),
  \urlprefix\url{http://link.aps.org/doi/10.1103/PhysRevE.60.6499}.

\bibitem[{\citenamefont{Toulemon et~al.}(2016)\citenamefont{Toulemon, Rastei,
  Schmool, Garitaonandia, Lezama, Catto{\"e}n, B{\'e}gin-Colin, and
  Pichon}}]{ToulemonEtal_AFM2016}
\bibinfo{author}{\bibfnamefont{D.}~\bibnamefont{Toulemon}},
  \bibinfo{author}{\bibfnamefont{M.~V.} \bibnamefont{Rastei}},
  \bibinfo{author}{\bibfnamefont{D.}~\bibnamefont{Schmool}},
  \bibinfo{author}{\bibfnamefont{J.~S.} \bibnamefont{Garitaonandia}},
  \bibinfo{author}{\bibfnamefont{L.}~\bibnamefont{Lezama}},
  \bibinfo{author}{\bibfnamefont{X.}~\bibnamefont{Catto{\"e}n}},
  \bibinfo{author}{\bibfnamefont{S.}~\bibnamefont{B{\'e}gin-Colin}},
  \bibnamefont{and} \bibinfo{author}{\bibfnamefont{B.~P.}
  \bibnamefont{Pichon}}, \bibinfo{journal}{Advanced Functional Materials}
  \textbf{\bibinfo{volume}{26}}, \bibinfo{pages}{2454} (\bibinfo{year}{2016}),
  \urlprefix\url{https://onlinelibrary.wiley.com/doi/abs/10.1002/adfm.201505086}.

\end{thebibliography}

\end{document}